\documentclass[prd,twocolumn,superscriptaddress,showpacs,aps,amssymb]{revtex4}
\usepackage{graphicx} % Include figure files
\usepackage{bm} % bold math
\newcommand{\beq}{\begin{equation}}
\newcommand{\eeq}{\end{equation}}
\newcommand{\bea}{\begin{eqnarray}}
\newcommand{\eea}{\end{eqnarray}}
\newcommand{\ba}{\begin{array}}
\newcommand{\ea}{\end{array}}
\newcommand{\lsim}   {\mathrel{\mathop{\kern 0pt \rlap
  {\raise.2ex\hbox{$<$}}}
  \lower.9ex\hbox{\kern-.190em $\sim$}}}
\newcommand{\gsim}   {\mathrel{\mathop{\kern 0pt \rlap
  {\raise.2ex\hbox{$>$}}}
\lower.9ex\hbox{\kern-.190em $\sim$}}}

\begin{document}

\title{Secondary protons from ultra high energy cosmic ray nuclei:
is the Greisen-Zatsepin-Kuzmin cutoff unavoidable?}
\author{R.~Aloisio}
\affiliation{INFN, Laboratori Nazionali del Gran Sasso, I-67010
  Assergi (AQ), Italy}
\author{V.~Berezinsky}
\affiliation{INFN, Laboratori Nazionali del Gran Sasso, I-67010
  Assergi (AQ), Italy}
\author{A.~Gazizov}
\affiliation{INFN, Laboratori Nazionali del Gran Sasso, I-67010
  Assergi (AQ), Italy}
%\affiliation{B.~I.~Stepanov Institute of Physics of the National
 % Academy of Sciences of Belarus,\\ 
 %BY-220072 Minsk, Belarus}

\begin{abstract}
We discuss production of ultra high energy secondary protons by 
cosmic ray primary nuclei propagating in intergalactic space through 
Cosmic Microwave Background (CMB) and infrared (IR) radiations. 
Under assumption that only primary 
nuclei with fixed atomic mass number $A_0$ are accelerated, the 
spectrum of
secondary protons is calculated. It is found that for all $A_0$ 
the diffuse flux of secondary protons starts to dominate over that of 
primary nuclei at energy $E \sim (1 - 2)\times 10^{19}$~eV, and thus 
the standard Greisen-Zatsepin-Kuzmin (GZK) cutoff is produced. 
%Uncertainties which can change this conclusion are discussed.    

\end{abstract}
      
\maketitle

%%%%%%%%%%%%%%%%%%%%%%%%%%%%%%%%%%%%%%%%%%%%%%%%%%%%%%%%%%%%%%%%%%%%%%%%
%%%{\em Introduction.---}%
%%%%%%%%%%%%%%%%%%%%%%%%%%%%%%%%%%%%%%%%%%%%%%%%%%%%%%%%%%%%%%%%%%%%%%

Ultra High Energy Cosmic Ray (UHECR) puzzle is one of the oldest 
puzzles in physics, which exists for more than 40 years and most probably 
now is close to being resolved. It appeared in 1966 together with 
prediction of the Greisen-Zatsepin-Kuzmin (GZK) cutoff \cite{GZK}. 
As physical phenomenon the GZK cutoff is explained by production of 
pions by UHE extragalactic protons interacting with CMB photons. In 
the diffuse spectrum the GZK feature (steepening of the spectrum)  
starts at energy $(3 - 5)\times 10^{19}$~eV. The UHECR puzzle was born 
simultaneously with prediction of the GZK feature, because already in 
1966 there were known at least three events with energies above the
GZK cutoff. With time the number of these events increased and in 
2004 there were known at least 16 events with energies higher than 
$1\times 10^{20}$~eV, much above the beginning of GZK cutoff. 
Now there are strong indications that steepening of the spectrum
compatible with GZK cutoff is observed by two detectors, HiRes 
and Pierre Auger Observatory (PAO). The observation of the GZK feature 
in differential spectrum measured by HiRes \cite{HiResGZK} is
confirmed by the measured value of $E_{1/2}=5.3\times 10^{19}$~eV 
in the integral spectrum. $E_{1/2}$ is a model-independent 
characteristic of the GZK cutoff in the integral spectrum 
\cite{BG88,BGGprd}, whose predicted value precisely coincides    
with value measured by HiRes. The Auger (PAO) differential spectrum 
is also consistent with the GZK steepening \cite{WatsonICRC}.

However, below the GZK cutoff the Auger data show the mixed mass 
composition, much heavier than pure proton composition 
\cite{WatsonICRC}.  If primary particles accelerated at the sources 
are heavy nuclei, why GZK cutoff, which is a signature of protons,  
is observed? This problem is further strengthened by the Auger observation  
of correlation with AGN at energy $E \gsim 6\times 10^{19}$~eV 
\cite{Auger-corr}, which also requires a proton-dominated composition. 
%Meanwhile in the Auger data there is inconsistency between mass
%composition measured at energy below the GZK cutoff  
%(mixed composition, considerably heavier than pure proton composition)
%and indication to almost pure proton composition at GZK cutoff 
%(observation of GZK cutoff, which is a proton feature and 
%correllation with with AGN at $E \gsim 6\times 10^{19}$~eV, which 
%requires the proton composition). 

In this paper we demonstrate that the pure nucleus 
primary component accelerated in sources results in the diffuse 
spectrum which is  strongly dominated by protons at energy above 
$(2 - 3)\times 10^{19}$~eV, i.e. at energy just below the beginning of
the GZK cutoff. This proton component is produced by 
photo-disintegration of primary nuclei on CMB.  

%%%%%%%%%%%%%%%%%%%%%%%%%%%%%%%%%%%%%%%%%%%%%%%%%%%%%%%%%%%%%%
%%%{\em Spectra of UHE nuclei and secondary protons. ---}%
We follow the paper \cite{ABG} in analytic calculations of spectra. 
We consider the expanding universe homogeneously filled by the sources of 
accelerated primary nuclei $A_0$ with  maximum energy    
$E_{\rm max}^{\rm acc}=Z_0\times 10^{21}$~eV, where $A_0$ and $Z_0$ are
atomic mass number and charge number, respectively. The generation
rate per unit comoving volume $Q_{A_0}(\Gamma,z)$ is given by  
\begin{equation}
Q_{A_0}(\Gamma,z)=\frac{(\gamma_g-2)}{m_N A_0}
{\mathcal L}_0 \Gamma^{-\gamma_g},
\label{eq:inj}
\end{equation}
where $z$ is redshift, $\Gamma$ is Lorentz factor of accelerated nucleus,
$\gamma_g$ is the generation index, $m_N$ is a mass of nucleon and 
${\cal L}_0$ is emissivity, i.e. energy generated per unit
comoving volume and per unit time at $z=0$. In Eq.~(\ref{eq:inj}) 
we assume $\Gamma_{\rm min} \sim 1$.  
%Limiting ourselves by
%conservative shock acceleration we consider $\gamma_g$ in the range 
%2.0 - 2.3. 

Propagating from a source a primary nucleus $A_0$ diminishes its 
Lorentz factor and atomic mass number due to interaction with background 
radiations, most notably with CMB and infrared (IR). The Lorentz factor
changes due to adiabatic ($d\Gamma/dt=-b_{\rm ad}(z)$) and 
$e^+e^-$ ($d\Gamma/dt=-b_{\rm pair}^A(\Gamma,z)$) energy losses 
and atomic number due to photo-disintegration losses 
($dA/dt=-b_{\rm dis}(A,\Gamma,z)$).   
An important characteristic of propagation is the mean time of nucleus 
photo-disintegration $\tau_A=1/b_{\rm dis}=(dA/dt)^{-1}$. 
\begin{figure*}[!t]
%\begin{minipage}[ht]{18cm}
\begin{center}
\includegraphics[width=0.9\textwidth]{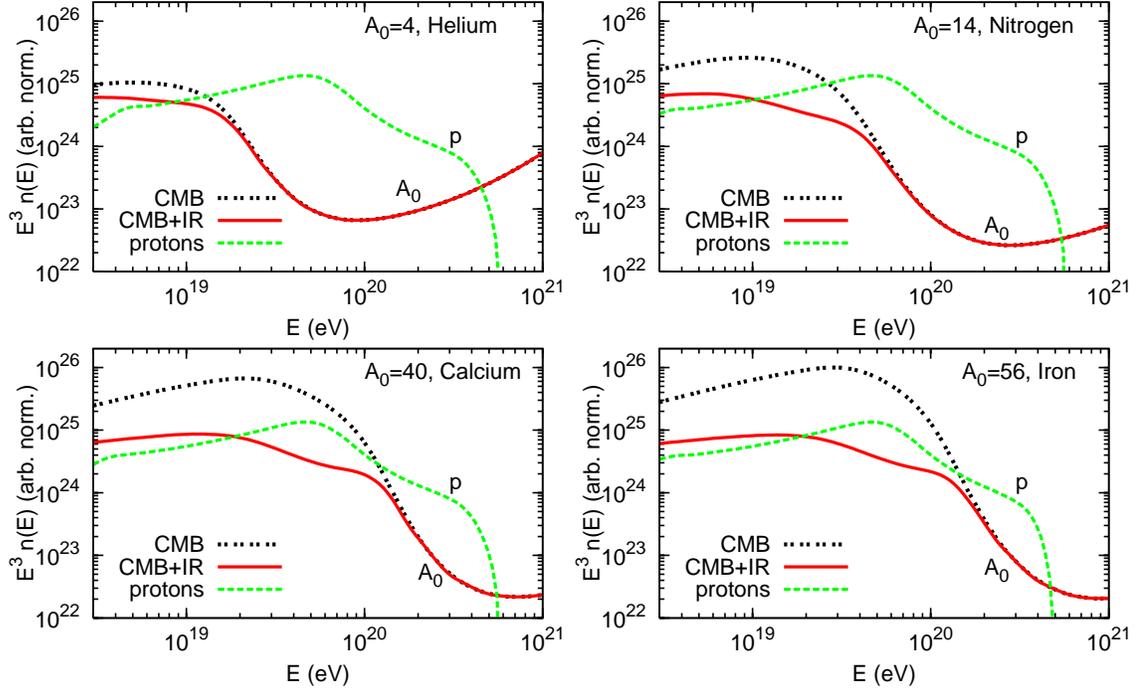}
\caption{Flux of primary nuclei $A_0$ interacting only with CMB  
(black dotted curves labelled by ``$A_0$'') and with interaction with IR 
included (red solid curves labelled by $A_0$) in comparison with 
secondary proton flux 
(green dashed curves labeled by ``p'') for different values of $A_0$
and for $\gamma_g=2.3$. 
In case of CMB+IR the secondary protons start to dominate at
energy $(1-2)\times 10^{19}$~eV.  
}
\label{fig:fluxALL}
\end{center}
%\end{minipage}
\end{figure*} 
\begin{figure*}[!t]
%\begin{minipage}[ht]{18cm}
\begin{center}
\includegraphics[width=0.9\textwidth]{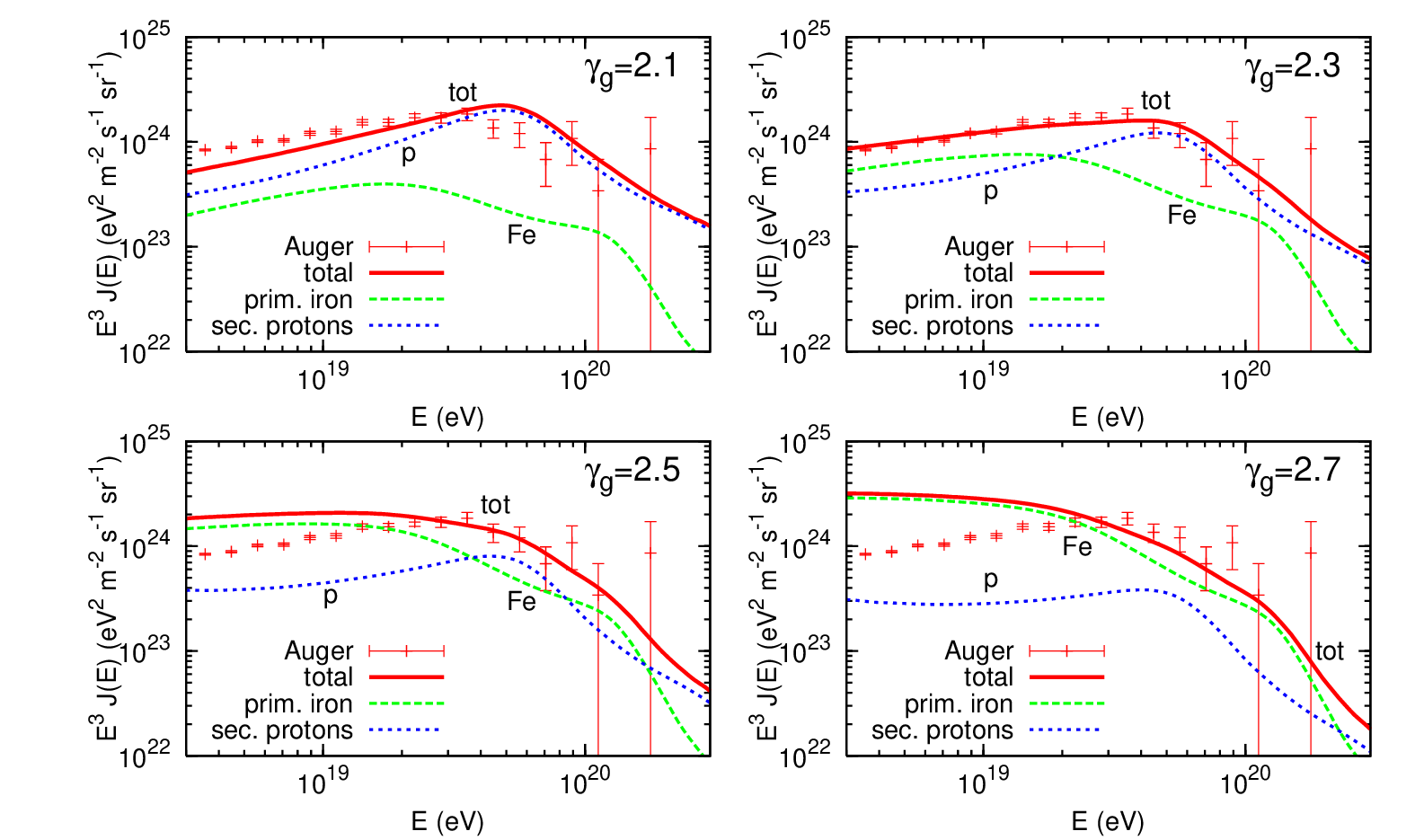}
\caption{Flux of primary iron nuclei (green dashed lines
labelled by ``Fe'')  and secondary protons (blue dashed lines
labelled by ``p'') for different generation indices $\gamma_g$.
The total spectrum (sum of primary iron and secondary protons) are 
shown by full red line and compared with the Auger surface detector 
spectrum. The energy of intersection of iron and secondary proton
spectra increases with $\gamma_g$. 
}
\label{fig:flux-gamma}
\end{center}
%\end{minipage}
\end{figure*} 
We study the nucleus evolution in the backward
time, whose role is played by redshift $z$. We consider as initial 
state the Lorentz 
factor $\Gamma$ at $z_0=0$  and calculate $\Gamma(z)$ for the fixed 
value $A$ using the equation   
\beq
 \frac{d\Gamma}{dz} = \left |\frac{dt}{dz} \right |
[b_{\rm ad}(z)+b^A_{\rm pair}(\Gamma,z)],
\label{eq:evol}
\eeq
where $dt/dz=-1/[(1+z)H(z)]$ and the Hubble parameter at 
redshift $z$ is $H(z)=H_0\sqrt{(1+z)^3\Omega_m+\Omega_{\Lambda}}$ with 
$\Omega_m$ and $\Omega_{\Lambda}$ being relative cosmological matter 
density and vacuum energy density, respectively. The numerical 
solution of Eq.~(\ref{eq:evol})
gives the evolution trajectory
\beq 
\Gamma_g(z)=G_A(\Gamma,z_0,z),
\label{eq:G}
\eeq
where the first two arguments describe the initial conditions.

The space density of species $a$ can be found from kinetic equation
\beq
\frac{\partial}{\partial t} n_a(\Gamma,t) - 
\frac{\partial }{\partial \Gamma}\left [ b_a(\Gamma,t) n_a(\Gamma,t) 
\right ] + \frac{n_a(\Gamma,t)}{\tau_a(\Gamma,t)} = Q_a(\Gamma,t).
\label{eq:kin1}
\eeq
The solution of Eq.~(\ref{eq:kin1}) for primary nuclei ($a=A_0$) 
accelerated at the source reads%
\begin{equation}
n_{A_0}(\Gamma)=\int_0^{\infty} dz_g \left | \frac{dt_g}{dz_g} 
\right | Q_{A_0}(\Gamma_g,z_g) 
\frac{d\Gamma_g}{d\Gamma} e^{-\eta(\Gamma_g,z_g)},
\label{eq:fluxA0}
\end{equation}
where $\Gamma_g$ is given by the evolution function (\ref{eq:G}), 
the generation function - by 
Eq. (\ref{eq:inj}) and analytic expression for $d\Gamma_g/d\Gamma$   
is given in \cite{ABG}. The quantity $\eta$ 
takes into account the photo-disintegration of the 
propagating nucleus: 
\beq
\eta(\Gamma_g,z_g)=\int_{t(z_g)}^{t_0}\frac{dt}{\tau_{A_0}(\Gamma(t),t)}~.
\label{eq:eta-t}
\eeq
The lower limit of integration in Eq.~(\ref{eq:fluxA0}) reflects the 
assumption of homogeneous 
distribution of the sources. The upper limit is imposed
by factor $\exp(-\eta)$ accompanied by condition of maximum
acceleration Lorentz factor: $Q_{A_0}(\Gamma_g)=0$,~ if 
$\Gamma_g \geq \Gamma_{\rm max}^{\rm acc}$. 

Secondary protons are produced in the photo-disintegration  chain of
the primary nucleus $A_0$ in the processes 
$\gamma +A \to (A-1) +N$, where $\gamma$ is a background photon and 
$N$ is a nucleon. The total flux of secondary protons is found 
in \cite{ABG} through the computation scheme in which 
the secondary protons from 
all intermediate nuclei with $A < A_0$ are taken into account. 
Another (approximate) method developed there is based on assumption 
of {\it instantaneous} decay (photo-disintegration) of primary nuclei 
to $A_0$ nucleons. Numerical computations have shown that 
fluxes calculated by both methods coincide with very good accuracy. 
We use here the more simple method of instantaneous decay.    

Every primary nucleus emitted by a source is considered as $A_0$ 
protons and the solution of Eq. (\ref{eq:kin1}) for $a=p$ 
with $\tau_p=\infty$ reads  
\begin{equation}
n^{\rm inst}_p(\Gamma)=A_0 \int_{z_g^{\rm min}(A_0)}^{z_g^{\rm max}}dz_g 
\left | \frac{dt_g}{dz_g} \right | Q_{A_0}(\Gamma_g^p,z_g)
\frac{d\Gamma_g^p}{d\Gamma}, 
\label{eq:flux-inst}
\end{equation}
where $\Gamma_g^p(z_g)$ is the Lorentz factor of a proton at generation,
i.e $\Gamma_g^p(z_g)=G_p(\Gamma,z_0,z_g)$, and ratio 
$d\Gamma_g^p/d\Gamma$ is given in \cite{BGGprd,ABG}. The term 
$e^{-\eta}$ in Eq.~(\ref{eq:flux-inst}) is absent because $\tau_p=\infty$.    
Since $Q_{A_0} \propto 1/A_0$  (see Eq. \ref{eq:inj}), the factor 
$A_0$ disappears from Eq.~(\ref{eq:flux-inst}), and the flux of
secondary protons is universal, i.e. the same for all $A_0$ \cite{ABG}.

The calculated diffuse spectra of primary nuclei and secondary protons are 
displayed in Fig.~\ref{fig:fluxALL} for different $A_0$, for $\gamma_g=2.3$ 
and for the
cases of CMB only and CMB+IR with IR/optical flux taken from \cite{Stecker06}.
The transition from primary nuclei to secondary protons occurs at energies   
$(1 - 2)\times 10^{19}$~eV for different $A_0$. The physics of this 
transition is based on universality of secondary-proton  spectrum.  
In the energy range displayed in Fig.~\ref{fig:fluxALL}, this spectrum is 
produced by photo-disintegration of nuclei on CMB, and 
the secondary-proton flux is the same with good accuracy in all four 
panels. The photo-disintegration of nuclei at energy 
$E_A \sim 10^{19}$~eV is dominated by IR/optical radiation, while the produced 
secondary protons have energies $E_p \sim E_A/A$, much below, in case 
of large $A$, the  discussed transition. Therefore, increasing the 
IR flux results in suppression of primary-nucleus spectrum without 
changing the secondary-proton spectrum at energies of interest.  
It lowers the energy of transition.   
\begin{figure*}[!t]
\begin{center}
\includegraphics[width=0.49\textwidth,angle=0]{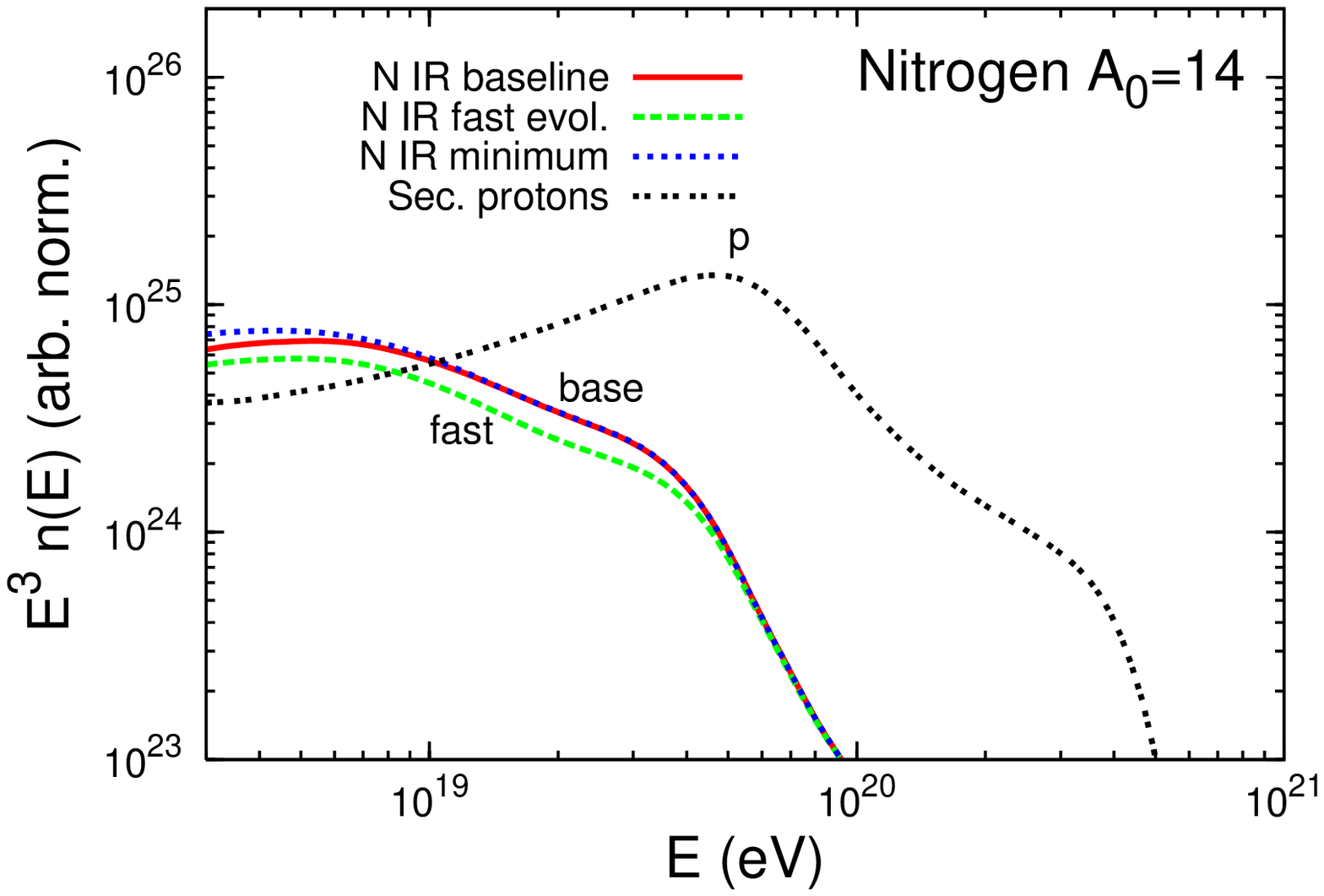}
\includegraphics[width=0.49\textwidth,angle=0]{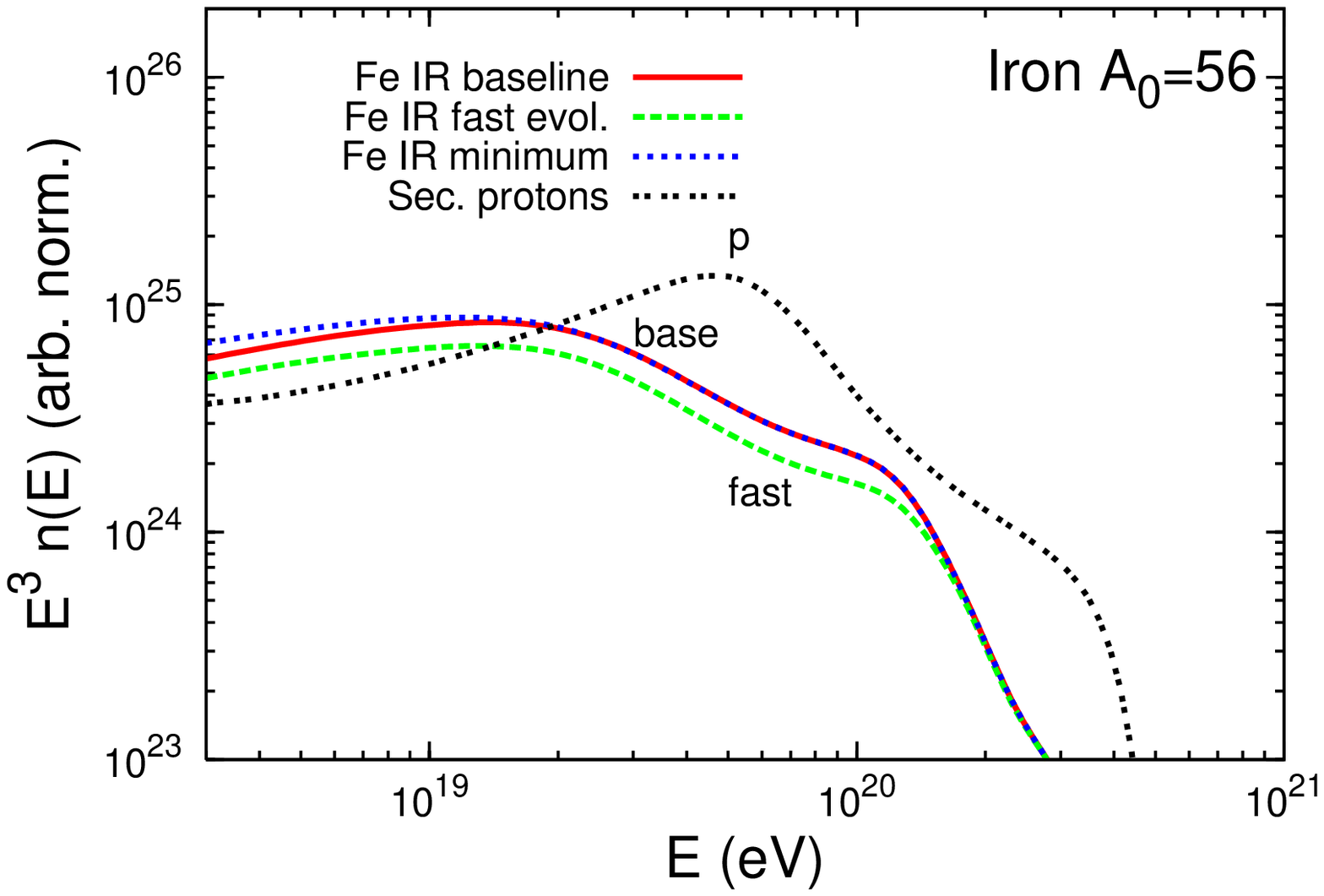}
\caption{The influence of IR/optical flux at epoch $z$ 
on energy of transition. 
The spectra of primary nuclei are displayed for three IR fluxes,
which are the same at $z=0$, but differ at $z>0$. 
The standard one \cite{Stecker06} shown by red solid curve labelled 
by ``base'', the larger flux with fast evolution \cite{Stecker06} 
is shown  by green dashed line and labelled by ``fast''and the
absolute minimum of the flux at redshift $z$ is shown by blue dotted 
line (it practically coincide with the baseline curve). The transition 
energy is given by intersection of proton flux (black dotted line) 
with one of the primary-nucleus curves. For minimum IR flux the energy of 
transition is the same as for the standard IR flux.
}
\label{fig:IRlim}
\end{center}
\end{figure*}
Fig.~\ref{fig:flux-gamma} demonstrates the dependence of transition
energy on generation index $\gamma_g$ and gives comparison of energy 
spectrum with Auger observations \cite{Auger-spectrum}. The
calculations are performed for 
four values of $\gamma_g$. One may observe that with increasing 
$\gamma_g$ from 2.1 to 2.7 the energy of transition increases. 
This effect is easy to understand. Let us consider the unmodified 
spectrum of primary nuclei $A_0$ and secondary protons in the   
toy model, where Lorentz factor is conserved, and photo-disintegration
to $A_0$ nucleons due to interaction with CMB is instantaneous.
The ratio of the secondary proton spectrum and unmodified nucleus
spectrum in terms of Lorentz factor is 
$n_p(\Gamma)/n_A^{\rm unm}(\Gamma)=A_0$. From this ratio it is easy to
obtain the ratio for equal energy fluxes 
$n_p(E)/n_A^{\rm unm}(E)=A_0^{2-\gamma_g}$. This ratio is suppressed
by large $\gamma_g$ and this suppression survives after the flux of 
primaries is reduced by photo-disintegration on IR radiation. This
mechanism predicts that dependence of transition energy on 
$\gamma_g$ is weaker for the light primaries and numerical calculations
confirm it.  

Fig.~\ref{fig:flux-gamma} shows that agreement with Auger spectrum is
much better for the flat generation spectra with $\gamma_g= 2.1 - 2.3$
than for the steep spectra. The transition energy for 
$\gamma_g \leq 2.3$ is less than $2\times 10^{19}$~eV.  

The generation spectra with $\gamma_g > 2.3$ are disfavored  by 
the observed spectrum, though it is allowed  theoretically in the 
shock acceleration \cite{KS}:
the distribution of the sources over maximum energy of acceleration 
$E_{\rm max}^{\rm acc}$ or
over luminosities makes the effective generation spectrum steeper at 
the highest energies.

Secondary-proton dominance depends on IR flux, which
suppresses the primary-nucleus flux. We shall demonstrate here that 
if IR flux measured at the present epoch $z=0$ is correct, the energy of 
transition remains the same or shifts (in case of strong evolution of
the sources) to the lower energies. We are interested thus in the minimum 
IR flux at $z >0$. We can prove the following general statement, valid
for any diffuse background radiation:\\ 
In the case of the generation rate of background radiation 
$Q(\epsilon,z)=K\epsilon^{-\alpha}(1+z)^m$, valid  up to $z_{\rm max}$, with 
arbitrary $\alpha$, $m \geq 0$ and $z_{\rm max}$, and under the 
assumption that 
the background photons are not absorbed, the flux of diffuse 
background radiation at epoch 
$z$ cannot be lower than $J_z(\epsilon)=(1+z)^{-3/2} J_0(\epsilon)$,
where $J_0(\epsilon)$ is the flux measured at $z=0$.

In Fig.~\ref{fig:IRlim} we illustrate the effect of variation of
the space density of IR photons $n_{\rm IR}(\epsilon,z)$ with 
redshift $z$. Three models of IR fluxes are used for calculations 
of the flux of UHE primary nuclei: the baseline model of 
Ref.~\cite{Stecker06}, which is considered as the standard model,    
the fast evolution model of Ref.~\cite{Stecker06}, which gives  the 
larger IR flux at higher $z$, and the absolute lower
limit for the IR flux at epoch $z$ as given above. 
%These cases are 
%shown in Fig.~(\ref{fig:IRlim}) by labels ``base'', ``fast'' and 
%``min'', respectively.  
The case of the  minimum 
IR flux is very close to the standard one and does not change our
conclusions. It is explained by the fact that the fluxes of UHE nuclei 
at $E > 1\times 10^{19}$~eV are produced at very small $z$ and thus
interact practically with $z=0$ flux of IR radiation, when the
source evolution is weak. \\*[2mm]
{\it In conclusion}, the preferential acceleration of heavy nuclei 
in the sources 
is naturally accompanied by dominance of the protons in the
diffuse spectrum at energy higher than $(1 - 2)\times 10^{19}$~eV 
with the standard GZK cutoff. We studied the extreme case of nuclei
acceleration solely, and found the dominance of secondary protons. 
In shock acceleration some fraction of accelerated protons is 
unavoidably present, and this primary component strengthens further 
the proton dominance \cite{Allard}. 
Another effect which increases the proton dominance is 
diffusion. The diffusion coefficient is inversely proportional to
the charge number $Z$ and thus time of nuclei propagation from 
nearby sources becomes longer and nuclei are destroyed. Variation 
of IR flux with $z$ also cannot increase the transition energy. 
The only effect which increases the transition energy is 
the large effective generation index of UHE nuclei $\gamma_g > 2.3$.
We conclude that the
GZK cutoff in the observational data is compatible with heavy mass
composition below the GZK cutoff in all most reasonable cases.\\*[2mm]
We are grateful to Floyd Stecker for helpful correspondence. 
This work is partially funded by the contract ASI-INAF I/088/06/0
for theoretical studies in High Energy Astrophysics.

\end{document}